# Dynamics of Clusters of Galaxies

*August E. Evrard* [*]

**Abstract**

The abundance and internal characteristics of rich clusters of galaxies can provide useful constraints on models of large–scale structure formation. The full power of observational constraints will only be realized when theoretical modeling of these complex, multi–component structures reaches a level of detail comparable to observations. This article will review some recent three dimensional, multi–fluid simulations of cluster dynamics and discuss their impact on issues raised from optical and X-ray observations of clusters. In 'bottom–up' formation scenarios (such as the ubiquitous cold dark matter model), galaxies form before rich clusters; hence, cluster formation is intimately linked to galaxy formation which, in turn, is tied to star formation. I will examine two issues which appear relatively insensitive to galaxy/star formation — the baryon fraction in clusters and the connection between X-ray morphology and $\Omega_o$ — and end with a topic that is intimately linked to it — dynamical biases in the cluster galaxy population.

## 1. Introduction

Clusters of galaxies are believed to be the largest collapsed, gravitationally bound structures in the universe. A typical rich cluster (Coma is the usual example) is a multi–component system containing hundreds of bright galaxies, a hot, metal–enriched intracluster medium (ICM) observed in X-rays, and dark matter whose presence has been inferred by application of the virial theorem for over 60 years [38].

Measures of the first two components, the galaxies and gas, are naturally linked by the star formation histories of the galaxies within the cluster. The present relative distributions of all three components reflect their full dynamical and thermal histories, which need not be the same. In particular, an exchange of energy between the different components will result in spatial segregation within the common cluster gravitational potential, with the coolest component being centrally concentrated and the hottest being extended.

Cluster formation must be understood in a cosmological context. Although a standard model of large–scale structure formation does not exist, many popular models envision clusters forming from gravitational amplification of small, initially Gaussian distributed density fluctuations [15, 13]. In nearly all viable models, cluster formation occurs after galaxy formation. Complete understanding of the present distributions of the dark matter, galaxies, and intracluster gas thus requires solving the FOE ('formation of everything') problem. That is, aspects of cluster formation are intimately tied to galaxy formation which, in turn, is linked to fragmentation of gas clouds and star formation [3]. Many unsettled issues regarding clusters (*e.g.*, the origin and distribution of metals in the ICM, whether or not galaxies fairly trace the cluster dark matter) persist because of the uncertainties in modeling galaxy/star formation.

---

[*]Department of Physics, University of Michigan, Ann Arbor, MI 48109-1120 USA. This work was supported by NASA Theory grant NAGW-2367 and NSF via supercomputer resources.

Despite this cautionary tone, there is good reason for optimism, since there are some important aspects of cluster formation which are largely decoupled from star/galaxy formation, at least in a model dependent fashion. In this article, I will briefly outline how one attempts to model cluster formation in a multi–component fashion using a combined N–body and smoothed particle hydrodynamics algorithm (§2). Most of the article is concentrated on applications of this approach to three particular issues — the baryon fraction in clusters (§3), the connection between cosmology and X–ray morphology (§4) and galactic dynamics within clusters (§5). The first two problems are fairly insensitive to galaxy formation while the last is strongly coupled to it.

## 2. SPH Modelling of Cluster Dynamics

Correctly modeling the dynamics of distinct components within a cluster requires a multi–fluid approach. The simulations discussed in this article all used a combination of an N–body algorithm to provide gravitational forces and smoothed particle hydrodynamics (SPH) [22, 16] to provide gas dynamic forces and thermal energy evolution for the gaseous baryonic component. Details of the P3MSPH code used in Sections 4 and 5 below can be found in ref [9]. Briefly, SPH is a Lagrangian scheme which uses a smoothing kernel $W(r,h)$ to determine characteristics of the fluid at a given point based on properties associated with the local particle distribution. For example, the density at the position of particle $i$ in the 'gather' interpretation [17] is given by

$$\rho_i = \sum_j m_j W(r_{ij}, h_i) \qquad (1)$$

where $r_{ij}$ is the separation of the pair of particles $i$ and $j$ and $h_i$ is the local smoothing scale. The kernel $W$ has compact support on a scale of a few $h$; hence, $h$ is a measure of the local resolution of the solution. Usually $h$ is adaptively varied both spatially and temporally such that a constant number of neighbors in the range $30 - 100$ is involved in the above sum. Applications to date have employed spherically symmetric kernels; schemes using anisotropic kernels are currently being developed [31].

The gas force on particle $i$ is found with a sum involving the gradient of the kernel

$$\left(\frac{\vec{\nabla} P}{\rho}\right)_i = \sum_j m_j \left(\frac{P_i}{\rho_i^2} + \frac{P_j}{\rho_j^2}\right) \vec{\nabla} W(r_{ij}, h_{ij}) \qquad (2)$$

where $h_{ij}$ is an average measure of $h_i$ and $h_j$ used to preserve pairwise symmetry in the force equations. This guarantees conservation of linear and angular momentum (for a spherically symmetric kernel) to machine accuracy. In regions of converging flow, an artificial viscosity is included to prevent free–streaming and increase entropy in a manner satisfying the local shock jump conditions. Radiative cooling terms can be included in the energy equation, as can heating terms due to, for example, photoionizing radiation or other sources.

The Lagrangian nature and wide dynamic range of SPH are well suited to the problem of large–scale structure formation. Schemes using Eulerian finite difference methods have recently come on line [7, 4]. A comparison of several cosmological gas dynamic schemes applied to structure formation in a cold dark matter universe has recently been done [18]. Examined at low resolution, the codes produce very similar results for the thermal and spatial structure of the gas. Higher resolution examination serves to illustrate the relative strengths and weaknesses of each approach. For low density contrasts, the Eulerian codes display superior resolution while at high density contrasts, higher resolution is achieved by the SPH codes. Roughly speaking, the 'break–even' density contrast $\delta_{eq}$, where Eulerian and Lagrangian approaches have comparable resolution, is that at which one particle in the SPH calculation is contained in one cell of the Eulerian code. At densities above $\delta_{eq}$, the Lagrangian method resolves one Eulerian cell with more than one particle (implying 'higher resolution') while for densities less than $\delta_{eq}$, a single Lagrangian particle covers many Eulerian cells ('lower resolution'). It follows then that $\delta_{eq} = N_{cell}/N_{part}$ where $N_{cell}$ is the number of cells in the Eulerian code and $N_{part}$ is the number of particles in the SPH code. In three dimensions, $\delta_{eq} = 256^3/64^3 = 64$ is a realistic value. Since the mass within an Abell radius of rich clusters represents a significant ($\delta > 100$) local density enhancement (see §3 below), a Lagrangian approach is (arguably, of course) currently the most efficient and effective means to model them numerically.

## 3.  The Cluster Baryon Fraction

The combination of X–ray surface brightness and spectral data for a cluster allows a direct estimate of the mass of hot, intracluster gas $M_{gas}$ to be made [30]. The mass in galaxies $M_{gal}$ can be estimated by multiplying the total optical luminosity in cluster galaxies by a representative, galactic mass–to–light ratio. Finally, the cluster binding mass $M_{tot}$ can be inferred by assuming the gas (and/or galaxies) are in hydrostatic equilibrium. From these masses, all determined within some radius $r_x$ where accurate X–ray and optical data exist, one can infer the mass fraction in *observed* baryons $f_b \equiv (M_{gal} + M_{gas})/M_{tot}$. It is interesting to compare this to the global value $<f_b> = \Omega_b/\Omega_o$ by defining the factor

$$\Upsilon \;=\; \left(\frac{\Omega_o}{\Omega_b}\right) \left(\frac{M_{gal} + M_{gas}}{M_{tot}}\right) \qquad (3)$$

where $\Omega_o$ is the present total density and $\Omega_b$ the global baryon density, each relative to the critical density.

This exercise has recently been carried out for the Coma cluster [36] and the results for the component masses within an Abell radius $r_A = 1.5\,h^{-1}$ Mpc ($h = {\rm H}_o/100$ km s$^{-1}$ Mpc$^{-1}$) are: $M_{gal} = 3.15 \pm 0.66 \times 10^{13}\,h^{-1}{\rm M}_\odot$, $M_{gas} = 5.66 \pm 1.02 \times 10^{13}\,h^{-5/2}{\rm M}_\odot$ and $M_{tot} = 1.10 \pm 0.18 \times 10^{15}\,h^{-1}{\rm M}_\odot$. The quoted $1\sigma$ errors are purely statistical, arising from uncertainty in optical photometry (for $M_{gal}$) and X–ray photon counts (for $M_{gas}$). In deriving the total mass, several independent estimates were derived which spanned the range 0.67 – $1.10 \times 10^{15} \, h^{-1} M_\odot$. The largest value was adopted, which was derived by scaling a set of 12 N–body/SPH simulations to the temperature of Coma and measuring the mass within $r_A$. The error is then based on the scatter among the runs. Adopting the largest value is conservative in that it leads to the minimum cluster baryon fraction. The total mass implies a mean overdensity within an Abell radius of $280/\Omega_o$.

The resultant baryon fraction in Coma is $f_b = 0.029 + 0.051 \, h^{-3/2}$ with a statistical uncertainty of about 25%. For $h = 0.5$, the baryons represent 17% of the total mass. The limits on the global baryon fraction from nucleosynthesis [32] are very stringent, $\Omega_b h^2 = 0.0125 \pm 0.0025$ at 95% confidence. The data then imply that the baryon fraction within an Abell radius in Coma is enhanced by the factor

$$\Upsilon \in [3.5 - 6.4] \, \Omega_o \qquad (4)$$

where the smaller factor assumes $h = 0.5$ and the larger $h = 1.0$.

From here, there are two possible avenues to pursue. One is to assume that the baryon fraction in Coma is representative of the global value ($\Upsilon = 1$) and so the data then constrain $\Omega_o$ to be in the range $0.16 - 0.29$. This perfectly valid line of reasoning is anathema to those cosmologists who prefer $\Omega_o$ to be unity either on the basis of aesthetics or because it is a natural outcome of inflationary models. (This may be a good time to point out that analysis of cluster X–ray morphology presented in the next section strongly supports this point of view.) The other avenue is thus to assume $\Omega = 1$, which leads to the following possibilities: (i) the estimated masses are systematically in error by large factors, (ii) the nucleosynthesis bound is in error or (iii) some process packed at least 3.5 times more baryonic matter than dark matter within an Abell radius in Coma. The last possibility may have physical justification in the fact that baryons can dissipate their thermal energy via radiative cooling while the dark matter may not. Classic cooling timescale arguments [37] suggest that dissipation should be modest on the large mass scales contained within an Abell radius. However, dissipation on galaxy scales may be very efficient, such that all the baryons in the universe are packed into very small structures before cluster formation. In that case, galaxies may infall nearly radially into the cluster center, and there deposit via inelastic collisions the baryons which they carried in, leading to a large baryon enhancement.

This possibility — that the baryon fraction in Coma is enhanced through 'known' mechanisms of gravity and dissipation — has recently been ruled out by a combination of semi–analytic and numerical analysis [36]. Treating the cluster as spherically symmetric, one can appeal to Bertschinger's self–similar infall solutions [1] to estimate the post–collapse structure for both dark matter and an 'infinitely dissipative' gas. The former is found by using the solution for infall of a collisionless gas and the latter comes from the solution which assumes accretion onto a central black hole. The solutions presented as a function of the self–similar radius have been re–phrased in terms of the mean interior overdensity $\Delta$ and are shown in Figure 1. At an overdensity of 280, an enhancement in the baryon fraction of $\sim 40\%$ is expected, well below the inferred range of equation (4). The cloud of points in Figure 1 are $[\Delta, \Upsilon]$ pairs constructed by random realizations of the quoted uncertainties in all the quantities on the right hand side of equation (3). The observations and model predictions are inconsistent at greater than 99% confidence.

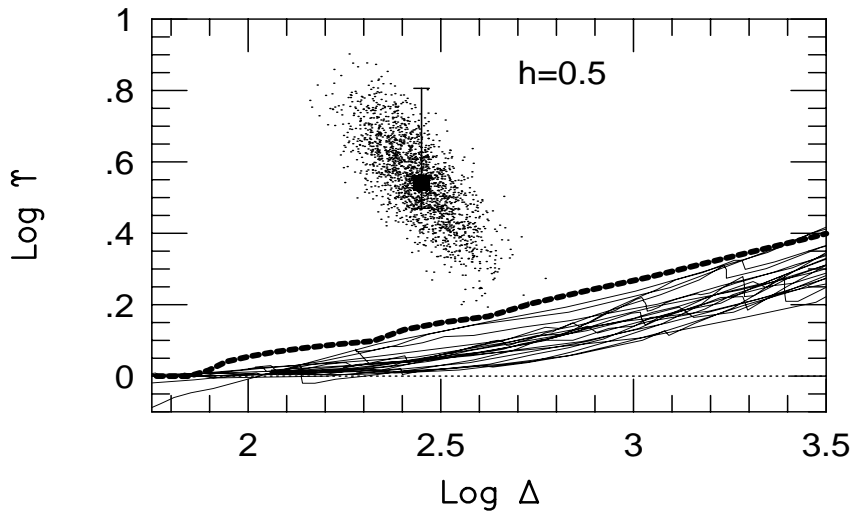

Figure 1. The baryon enhancement factor $\Upsilon$ as a function of mean interior overdensity for the spherical, self–similar accretion model (thick, dashed line) and the results of $N$-body/SPH simulations of 'pressureless' baryon accretion in a CDM universe. The dots represent $(\Upsilon, \Delta)$ pairs drawn at random using the uncertainties for $\Omega_b$ and for the masses of the different components assuming $h = 0.5$. Arrows show how the value of $\Upsilon$ would change for $h$ in the range of 0.4 to 1.0. (from White *et al.* 1993.)

To verify that the spherical model results are relevant to more realistic, fully three dimensional accretion, a series of N–body/SPH runs were performed using CDM cluster initial conditions. The code used was a combined TREE/SPH code written by J. Navarro [26]. To mimic a perfectly dissipative gas, an isothermal equation of state was assumed, using a temperature well below that of any resolvable potential well in the simulation. A set of runs, each with 65536 particles representing gas and dark matter, were performed. The enhancements seen in the simulations, shown as the light lines in Figure 1, all fall below the spherical model solution. That the spherical model should provide an upper limit to the enhancement factor is perhaps not too surprising, since it represents the limiting case of zero angular momentum for all mass elements. In the three dimensional case, sublumps acquire angular momentum and may 'miss' the center on first infall, resulting in a more extended baryon distribution.

These toy models produce clusters which are far from realistic. The models contain no hot, intracluster medium while observations indicate that the majority of baryons end up in this phase. A more realistic treatment, such as the models discussed in the next section, show the ICM to be slightly more *extended* than the underlying dark matter, *i.e.*, $\Upsilon < 1$. This result is due energy exchange between the dark matter and ICM during mergers [27, 28].

The upshot is that a large (factor $>3$) enhancement in the baryon fraction within an Abell radius in a rich cluster like Coma is impossible with conventional dynamics. Consistency with $\Omega_o = 1$ requires other explanations for the discrepancy, such as those listed above.

## 4. A Morphology–Cosmology Connection

There are several ways clusters can be used as cosmological diagnostics. Their abundance as a function of, for example, velocity dispersion $\sigma$, is extremely sensitive to the normalization of the fluctuation spectrum [35]. However, small errors in $\sigma$ can translate into large errors in the normalization [10], and handling this correctly requires both good data and careful analysis. Because of degeneracy between the spectrum normalization and $\Omega_o$, abundances provide little information on $\Omega_o$ (unless the spectrum normalization is known by other means).

A more fruitful approach to constraining $\Omega_o$ is to examine the structure of the hot, intracluster gas. The motivating idea is that, because the linear growth of perturbations diminishes as $\Omega$ decreases, structure formation in a low $\Omega_o$ universe should occur earlier than if $\Omega = 1$. An analysis based on a spherical model for cluster collapse yields an age difference between clusters in models with $\Omega_o = 0.2$ and $\Omega = 1$ of $\sim 0.3$ $H_o^{-1} \sim 4 - 6$ billion years [29]. The sound crossing time for 10 keV gas in the central 1 Mpc of a cluster is only 0.6 billion years, so this age difference corresponds to many sound crossing times within the region surveyed by X–ray imaging instruments. This leads to the expectation that clusters in low density models should have more relaxed X–ray isophotes than their critical counterparts.

This effect has now been verified and quantified with a set of 24 P3MSPH simulations [11]. Eight sets of initial conditions, two each in comoving periodic boxes of side 30, 40, 50 and 60 Mpc ($h = 0.5$), were generated in a constrained manner [2] from an initial CDM fluctuation spectrum. Each initial density field was evolved in three different cosmological backgrounds: (i) an unbiased, open universe with $\Omega_o = 0.2$; (ii) an unbiased, vacuum energy dominated universe with $\Omega_o = 0.2$ and $\lambda_o = 0.8$ and (iii) a biased, critical density ($\Omega = 1$) universe with *rms* present, linear mass fluctuations in a sphere of $8\,h^{-1}$ Mpc equal to $\sigma_8 = 0.59$. A baryon content of $\Omega_b = 0.1$ was assumed for the models, with all the baryons in the form of gas. The rest of the mass was assumed to be collisionless dark matter.

Present day X–ray images of half of the simulated clusters are shown in Figure 2, along with a set of *Einstein* IPC images of four Abell clusters. The simulated images show the IPC band–limited flux in X–rays with an angular resolution of about $1'$. They are cleaner than the observations because no noise or background of points sources have been added. The low density models, shown in the first two rows, are much more centrally concentrated and display much less asymmetry than the critical universe clusters. These differences arise because the low $\Omega_o$ clusters suffer fewer merging events at late times, a result expected from analytic arguments [29, 21, 19]. We have quantified the differences using statistics measuring the surface brightness fall-off (the familiar $\beta_{fit}$ parameter), mean isophotal center shift [25] and the mean eccentricity. The same measures have been made for both the simulated and observed clusters. Histograms of any of these show a distinct difference between the low and high density models, with the observations strongly favoring $\Omega = 1$ over either of the $\Omega_o = 0.2$ universes [24]. This result is supported by recent analysis of the abundance of rich clusters. In order to reproduce observations, an $\Omega_o = 0.2$ CDM dominated universe requires a very high fluctuation amplitude $\sigma_8 = 1.25 - 1.58$, which requires galaxies to be *less* clustered than the mass distribution [35].

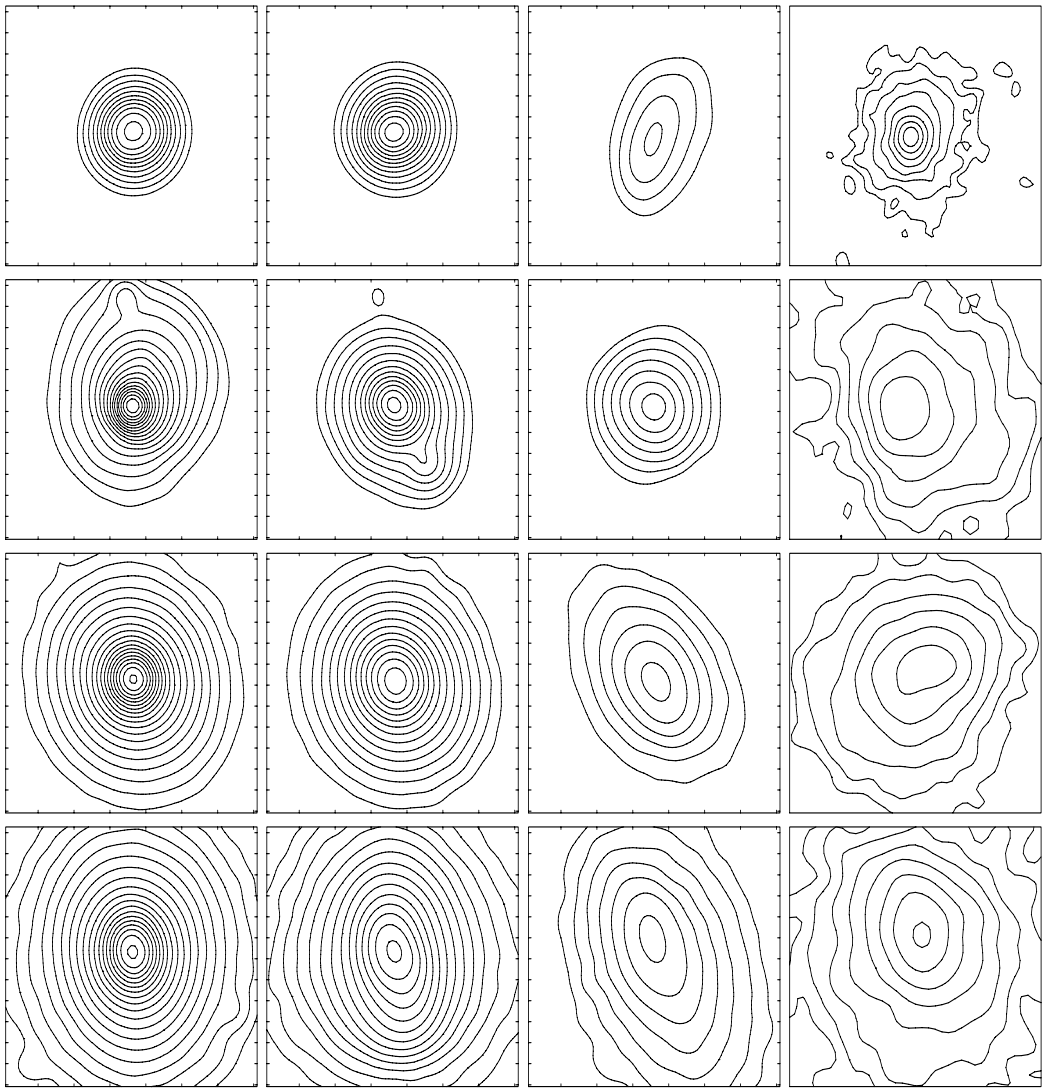

Figure 2. Contour maps of cluster X-ray emission showing the dependence of X-ray morphology on the underlying cosmology. The first three columns show simulated clusters 'viewed' at $z = 0.04$ evolved in different cosmologies (from left to right): (i) $\Omega_o = 0.2$; (ii) $\Omega_o = 0.2$, $\lambda_o = 0.8$; and (iii) $\Omega = 1$, $\sigma_8 = 0.59$. The fourth column consists of *Einstein* IPC observations of Abell clusters; from top to bottom, they are: A496, A399, A2256 and A401. Each row in the first three columns corresponds to the same initial density field generated in (from top to bottom) 30, 40, 50 and 60 Mpc cubes. The contours in all maps are spaced by factors of 1.8 in surface brightness from $3.5 \times 10^{-4}$ cts s$^{-1}$ arcmin$^{-2}$, and the spatial scale for every map is identical, the distance between tic-marks being 188 kpc. (from Evrard *et al.* 1993.)

There is a potential conflict between this result and that of the previous section. The large baryon fraction in Coma could be explained by a low $\Omega_o$, but the morphology of X–ray emission from clusters strongly disfavors low values. One possibility is that $\Omega_o \sim 0.6$, and both constraints are satisfied. (This idea can be discounted by appealing to the Principle of Non–Ugliness.) One might instead doubt the robustness of the X–ray morphology constraints. After all, the physics in the models shown in Figure 2 is fairly simplistic, incorporating gravity for both components, as well as shock heating and a thermal pressure gradient for the gas.

Although there is more physics beyond this simple approach, it is difficult to finger a mechanism which would strongly distort in an *anisotropic fashion* the present cluster X–ray morphologies in the low $\Omega_o$ models. Feedback due to winds from early–type galaxies would have to occur very recently, and the winds would have to be coherently directed so as to distort the isophotes in a manner similar to that which occurs naturally by merging. Recent simulations incorporating winds in $\Omega = 1$ clusters show little effect on the overall morphology [23]. Adding radiative cooling would produce a large central cooling flow, but there is no reason to suspect this will strongly affect the morphology of the outer regions. Finally, unrealistically large tidal torques would be required to distort the structure of the X–ray gas in the inner $\sim 1$ Mpc region, where the bulk of the X–rays are observed.

It all boils down to this. Generating anisotropy in the gas distribution at late times requires a directed source of energy input with magnitude comparable to the binding energy of the cluster. The most natural source for such directed energy is the merging of two systems of roughly comparable mass. To save the low $\Omega_o$ models, one needs to come up with a mechanism(s) which replaces merging, but produces the same effects. It is not at all clear how to do this.

## 5.   Galactic Dynamics in Clusters

As a final topic, let's turn from X–ray to optical wavelengths. The theme common with the preceding two sections is the determination of $\Omega_o$ from clusters. It is well known that dynamical mass estimates based on the virial theorem have yielded mass to light ratios around $150\,h^{-1}$ in solar units, which is about a factor 5 smaller than that required to reach closure density [12, 14]. Again, the two possible interpretations are: (i) the estimate is unbiased and $\Omega_o \sim 0.2$ or (ii) there is a systematic bias in the estimate which makes it consistent with $\Omega = 1$. One possibility for the latter is that the dark matter is very weakly clustered on comoving scales $\lesssim 10$ Mpc. Another is that the galaxies are condensed toward the cluster center, and that one is measuring only some inner fraction of the total cluster mass and missing an extended, outer dark envelope. The latter issue has been investigated numerous times with N–body experiments over the past decade. Unfortunately, the interpretation of these simulations is clouded by the rather naive way in which galaxies were represented within the cluster. In some studies, heavy particles meant to represent the luminous parts of galaxies were merely put in 'by hand' in the initial conditions [8, 33]. Others tagged a subset of the collisionless dark matter particles, based on plausible physical arguments, to represent the kinematics of the galaxy population [5].

These experiments generally produced results in the desired direction; that is, the galaxies represented a cooler, condensed population within the dominant dark matter potential well and mass estimates based on them underestimated the total mass of the system. However, the magnitude of the effect varied by rather large factors, depending on the treatment. Furthermore, the physics responsible for the bias remains poorly understood. Dynamical friction [6], incomplete or 'non–violent' relaxation [34, 39], or some combination of the two remain the most viable mechanisms.

The limiting factor in performing such experiments is the uncertainty involved in galaxy formation. Ideally, one would like to form galaxies *in situ* and subsequently follow their hierarchically clustering to the scale of rich clusters. The combined N–body/gas dynamic methods are now making this possible. In the simplest scenario, one allows the baryons to radiatively cool within their parent dark halos. Since local temperatures and densities are known, cooling rates can be calculated. The principle uncertainty is due to lack of resolution — the baryons represented by a single particle or single cell, which is typically $\gtrsim 10^8 M_\odot$, is assumed to be a single phase medium characterised by the given density and temperature. This treatment, though not perfect, has much more physical validity than tagging particles in an N–body experiment. The end result is a two–phase structure in the baryons, with cold, dense knots one associates with galaxies surrounded by halos of hot, rarefied gas.

Simulations of this sort have been successful recently in producing clusters with anywhere from three [20] to several tens [12] of such 'galaxies' within them. An example is shown in Figure 3. This cluster was modeled with P3MSPH using $2 \times 64^3$ particles to represent the dark matter and baryons. The simulation modeled a periodic cube 22.5 Mpc on a side ($h = 0.5$). The limiting spatial resolution was $\sim 30$ kpc and the mass per baryon particle was $3 \times 10^8 M_\odot$. An $L_*$ galaxy would thus be modeled by about 300 particles. Radiative cooling for the baryons based on collisional ionization equilibrium was included. One of the main aims of such a simulation is to address the issue of dynamical biases in the galaxy population.

The results are striking. The galaxies, shown in the middle panel of Figure 3, are much more centrally concentrated than the dark matter. Table 1 gives a summary of the relevant properties. Galaxies were defined to be cluster members if they lie within a radius of 1.6 Mpc from the cluster center. This radius is that within which the mean density, based on the known dark mass distributions, is 180 times the background value. The number of galaxies so found $N_{gal}$ is shown for galaxies with particle counts above $N_{cut}=32$ and 128.

Table 1 : Cluster Parameters from Pure SPH Treatment

| $N_{cut}$ | $N_{gal}$ | $\sigma_{gal}$ ( km s$^{-1}$) | $\sigma_{gal}/\sigma_{DM}$ | $R_{gal}/R_{DM}$ | $M_{vir}/M_{true}$ |
|---|---|---|---|---|---|
| 32 | 29 | 458 | 0.84 | 0.35 | 0.43 |
| 128 | 10 | 469 | 0.86 | 0.22 | 0.25 |

The galaxies represent a cooler population, as witnessed by the values of the 'velocity bias' parameter, the ratio $\sigma_{gal}/\sigma_{DM}$ where $\sigma$ is the one–dimensional galaxy velocity dispersion obtained from an average of the three orthogonal directions. A modest $\sim 15\%$ bias is evident in the velocities. In contrast, the ratio of the half–mass radii $R_{gal}/R_{DM}$ (determined from the known three dimensional positions and using 1.6 Mpc as an outer radius) shows a much more pronounced bias. Galaxies with baryon mass above $10^{10} M_\odot$ (32 particles) are more concentrated than the dark matter by a factor 3, while galaxies above a mass cut a factor 4 larger are even more concentrated. Application of the virial theorem to determine binding masses results in a large (factor $2-4$) underestimate of the total mass of the cluster.

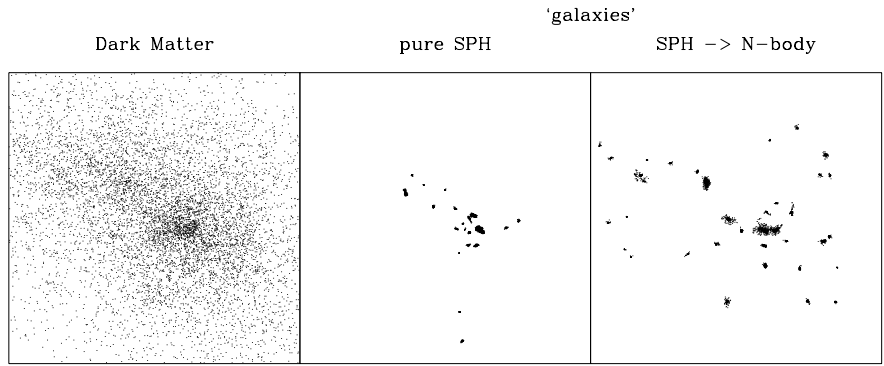

Figure 3. Distribution of particles in a 1 Mpc region centered on a rich cluster at $z=0$ from a P3MSPH simulation using CDM initial conditions. The left panel shows a random subset of the dark matter particles (the total number in the cluster is nearly 56,000). The middle panel shows baryonic particles found in 'galaxies' at the end of the simulation, using SPH dynamics throughout the evolution. The right panel shows the final 'galaxy' distribution when particles in galaxies are treated as collisionless 'stars' from $z=0.7$ to $z=0$. All galaxies with more than 32 particles are shown. See the text for further discussion.

What is worrisome about this treatment is that the galaxies are assumed to be purely gaseous throughout the evolution of the cluster. Their interactions with the surrounding medium and with each other during collisions entail viscous drag, which is unphysical for a galaxy comprised mainly of stars. Of particular concern is the fact that the largest galaxy in the center of the cluster ends up containing *more than half* of the total baryons in cluster galaxies. Although bright, central cD galaxies are not uncommon in rich clusters, it is not the norm for the central cD to be brighter than the sum of all the other galaxies in the cluster.

A simple way to test the effects of this collisional, or 'pure SPH', treatment on the galactic dynamics within the cluster is to turn the SPH gas particles in galaxies into collisionless 'stars'. Ideally, the star formation process should be modeled self–consistently within the code; however, there still exists a large amount of uncertainty in parameterising star formation rates and the associated feedback. For the purposes of examining galactic orbits within the cluster, the key element is that the star formation in galaxies occur before the bulk of the cluster is assembled. With this in mind, the following simple experiment was performed as a variation to the original SPH run. At a redshift $z=0.7$, before cluster collapse but after many of the cluster galaxies were assembled, the particles labeled in galaxies were 'instantaneously' turned into collisionless 'star' particles. At this time, all the remaining gas was removed and the mass associated with it was added to the dark matter particles. A collisionless, two–fluid run, using as initial conditions the dark matter particles and galaxies comprised of the star particles above, was then evolved from $z = 0.7$ to $z = 0$, a time interval of 7.3 Gyr with $h = 0.5$. This run thus follows the evolution of the same cluster as in the pure SPH run, but with the galaxies being treated as collisionless entities during the epoch of cluster formation. This 'SPH $\rightarrow$ N-body' treatment is arguably more realistic, if one believes the observed stellar populations in cluster galaxies pre–date the cluster itself.

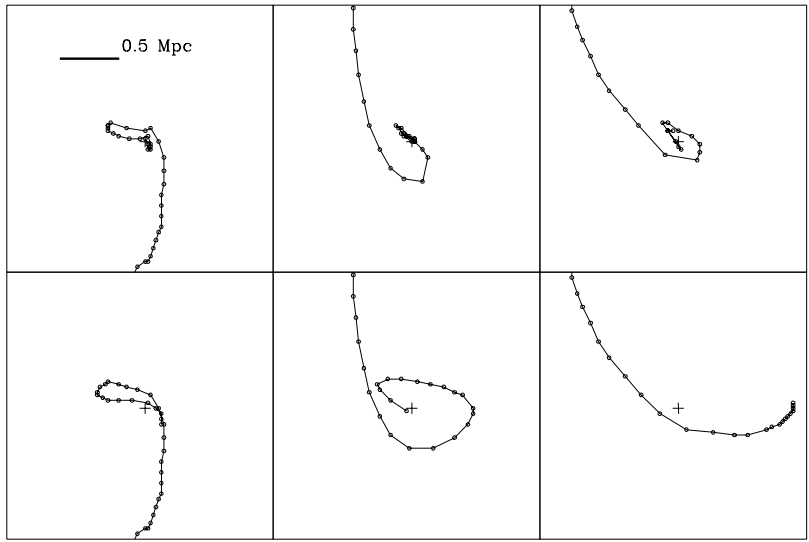

Figure 4. Orbits for 'galaxies' in the same cluster when treated as collisional (top row) or collisionless (bottom row) entities. Each row shows the orbit of the same object identified at $z = 0.7$, the starting epoch for the collisionless run. From left to right are shown the 2nd, 12th and 14th most massive galaxies identified at that time. The cross marks the center of the cluster at the final epoch. Dots sample the orbits at intervals of roughly $2 \times 10^8$ yr, ending at $z = 0$. The orbits are clearly sensitive to the dynamical treatment. In the pure SPH treatment, galaxies tend to hit the center and 'stick', whereas their collisionless counterparts sail through relatively unperturbed.

The resulting cluster galaxy distribution is shown as the right panel in Figure 3. The galaxies in the SPH $\rightarrow$ N-body treatment are clearly more extended than their pure SPH counterparts. Figure 4 shows the orbits of three galaxies identified at $z = 0.7$. The top row shows the orbit under the pure SPH treatment while the lower row shows orbits under the SPH $\rightarrow$ N-body treatment. The predominantly radial infalling orbits take the galaxies very close to the cluster center. In the SPH case, the viscous gas interactions in the high density, central region prove effective at 'braking' the galaxies. This prevents them from completing a full orbit and also enhances the accretion rate onto the large, central galaxy. In contrast, the galaxies comprised of collisionless stars fly through the center relatively undisturbed, as expected if the cluster velocity dispersion is larger than the internal velocities of the galaxies (which is the case here). No extremely large central galaxy forms. Instead, two galaxies separated by 0.5 Mpc, each roughly one–fifth the mass of the largest in the pure SPH run, are the most conspicuous objects in the cluster at $z = 0$.

A listing of the relevant cluster properties inferred from the collisionless treatment is given in Table 2. Because of the reduced merger rate, the number of galaxies in the cluster at $z=0$ nearly doubles over the pure SPH case. A modest velocity bias persists, being slightly larger for the more massive galaxies. The spatial distribution depends on the galaxy mass cutoff. The massive subsample is more concentrated than the dark matter while the set of all galaxies above the minimum 32 particle count ($10^{10} M_\odot$ in baryons) is spatially unbiased with respect to the dark matter. The resulting virial mass estimates reflect this trend; the massive subsample underestimates the total mass by a factor of three, but analysis based on the full sample produces an estimate accurate to within 10%. The different behavior of the two mass groups is not a transient result, since the same trend exists at earlier redshifts. It may be that the result is 'inherited' from the SPH run via the initial conditions; the gravitational clustering being inefficient at erasing the memory of the initial bias.

Table 2 : Cluster Parameters from SPH → N-body Treatment

| $N_{cut}$ | $N_{gal}$ | $\sigma_{gal}$ ( km s$^{-1}$) | $\sigma_{gal}/\sigma_{DM}$ | $R_{gal}/R_{DM}$ | $M_{vir}/M_{true}$ |
|---|---|---|---|---|---|
| 32 | 52 | 479 | 0.89 | 1.08 | 0.92 |
| 128 | 17 | 386 | 0.71 | 0.56 | 0.32 |

To summarize, the issue of biases in the galaxy distribution within clusters remains uncertain. There appears to be a growing trend toward modest amounts of velocity bias, $\sigma_{gal}/\sigma_{DM} \sim 0.7 - 0.9$, but less clear ideas on the relative spatial distributions of galaxies and dark matter. The problem is complicated by the fact that equilibrium models are not appropriate for clusters formed in flat cosmologies, since they will generally have experienced considerable merging on a timescale comparable to their dynamical time.

## 6.  Epilogue

Dynamical modeling of clusters of galaxies has improved significantly in the past few years, with the advent of simulation algorithms capable of handling the coupled evolution of multiple components representing dark matter, intracluster gas and galaxies. The new generation of experiments has yielded insight into the problem of the baryon fraction in clusters, and has provided details on the connection between cosmology and the X–ray morphology of clusters. The latter is a prime example of the type of problem which would be virtually impossible to tackle with a pure N–body approach.

Issues which are more intimately linked to galaxy/star formation (the FOE problem) remain relatively poorly understood. Although definitive answers to the question of dynamical biases for galaxies in clusters remain elusive, results emerging from a variety of independent treatments suggest that galaxies should give a velocity dispersion estimate biased slightly $(10 - 30\%)$ low with respect to the dark matter. Optical mass estimates are likely to underestimate the total binding mass, but the magnitude of this effect is fairly uncertain.

At this point, clusters appear schizophrenic regarding the value of $\Omega_o$. Their X–ray morphology and abundance prefer $\Omega = 1$ over low $\Omega_o$, but the baryon fraction favors the opposite.

Their mass–to–light ratios could point in either direction, depending on the degree of dynamical biasing. This last issue will likely be settled within the next few years by very high resolution simulations incorporating star formation in a self–consistent fashion. In short, the deep waters of the non–linear evolution of clusters remain murky, but the edges of the pool, at least, are beginning to clear.

I would like to thank my collaborators in the above projects — J. Mohr, D. Fabricant, M. Geller, S.D.M. White, J. Navarro, C. Frenk, F. Summers and M. Davis — for allowing me to use our joint results in this article.